%% file: 0_main_single_column.tex
\documentclass[
	reprint,
	amsmath,amssymb,
	superscriptaddress,
	twocolumn,
	accepted=2021-01-17
   ]{quantumarticle}

\newcommand{\Tr}{\text{Tr}}

\makeatletter
\newsavebox{\@brx}
\newcommand{\llangle}[1][]{\savebox{\@brx}{\(\m@th{#1\langle}\)}%
  \mathopen{\copy\@brx\kern-0.5\wd\@brx\usebox{\@brx}}}
\newcommand{\rrangle}[1][]{\savebox{\@brx}{\(\m@th{#1\rangle}\)}%
  \mathclose{\copy\@brx\kern-0.5\wd\@brx\usebox{\@brx}}}
\makeatother

\usepackage{ulem}

\newcommand{\kket}[1]{
	\ensuremath{|{#1}\rrangle}
}
\newcommand{\bbra}[1]{
	\ensuremath{\llangle{#1}|}
}
\newcommand{\bbraket}[1]{
	\ensuremath{\llangle{#1}\rrangle}
}
\newcommand{\red}[1]{
	{#1}
}

\newtheorem{lemma}{Lemma}

\usepackage[numbers,sort&compress,square]{natbib}
\usepackage{graphicx}
\usepackage{color}
\usepackage{dcolumn}
\usepackage{siunitx}
\usepackage{bm}
\usepackage{braket}
\usepackage[utf8]{inputenc}

\begin{document}


\title[]{Overhead for simulating a non-local channel with local channels by quasiprobability sampling}

\author{Kosuke Mitarai}
\email{mitarai@qc.ee.es.osaka-u.ac.jp}
\affiliation{Graduate School of Engineering Science, Osaka University, 1-3 Machikaneyama, Toyonaka, Osaka 560-8531, Japan.}
\affiliation{Center for Quantum Information and Quantum Biology, Institute for Open and Transdisciplinary Research Initiatives, Osaka University, Japan.}
\affiliation{JST, PRESTO, 4-1-8 Honcho, Kawaguchi, Saitama 332-0012, Japan.}

\author{Keisuke Fujii}
\affiliation{Graduate School of Engineering Science, Osaka University, 1-3 Machikaneyama, Toyonaka, Osaka 560-8531, Japan.}
\affiliation{Center for Quantum Information and Quantum Biology, Institute for Open and Transdisciplinary Research Initiatives, Osaka University, Japan.}
\affiliation{Center for Emergent Matter Science, RIKEN, Wako Saitama 351-0198, Japan}

\date{\today}

\begin{abstract}
    \input{01_abstract.tex}
\end{abstract}

\maketitle

\input{10_introduction.tex}

\input{30_method.tex}

\input{40_other_approaches.tex}

\input{70_conclusion.tex}

\appendix
\input{80_appendix.tex}

\end{document}

%% file: 01_abstract.tex
As the hardware technology for quantum computing advances, its possible applications are actively searched and developed.
However, such applications still suffer from the noise on quantum devices, in particular when using two-qubit gates whose fidelity is relatively low.
One way to overcome this difficulty is to substitute such non-local operations by local ones.
Such substitution can be performed by decomposing a non-local channel into a linear combination of local channels and simulating the original channel with a quasiprobability-based method.
In this work, we first define a quantity that we call channel robustness of non-locality, which quantifies the cost for the decomposition.
While this quantity is challenging to calculate for a general non-local channel, we give an upper bound for a general two-qubit unitary channel by providing an explicit decomposition.
The decomposition is obtained by generalizing our previous work whose application has been restricted to a certain form of two-qubit unitary.
This work develops a framework for a resource reduction suitable for first-generation quantum devices.

%% file: 10_introduction.tex
\section{Introduction}

We now have a programmable quantum device whose dynamics cannot be simulated by a classical computer within its runtime \cite{Arute2019}.
However, the capability of such devices is rather limited because of the absence of the quantum error correction.
They are frequently referred to as noisy intermediate scale quantum (NISQ) devices \cite{Preskill2018quantumcomputingin}.
There has been a substantial amount of research efforts to develop useful applications of NISQ devices in recent years \cite{Peruzzo2014, RevModPhys.92.015003, Farhi2014,Mitarai2018, farhi2018classification,bravoprieto2019variational,LaRose_2019}.
The weakness of NISQ devices is that the number of qubits, the fidelities of gates, and the connectivity are limited.
The gate fidelities are especially restricted for two-qubit entangling gates.
One approach to circumvent such limitation is to use so-called variational quantum algorithms.
\red{They employ parametrized quantum circuits and optimize the parameters to perform a given task.
In such algorithms, we frequently construct the largest possible circuit allowed on a device to maximize the advantage of the use of quantum devices.}

While this approach is promising as it can in principle employ such circuits, such algorithms can still be improved if one can perform further resource reduction.
For example, if we can reduce the number of qubits or two-qubit gates required to obtain an output from a certain quantum circuit, it would widen the range of circuits that can be used for variational algorithms.
To this end, a few approaches have been proposed.
One is to decompose a large circuit into smaller ones by ``cutting'' circuits using a tomography-like method \cite{peng2019simulating}.
Also, in Ref. \cite{mitarai2019constructing}, we have presented a method to ``cut'' a certain non-local gate by decomposing it into a linear combination of local operations.
These approaches share the same property that the overhead for the decomposition, which in this context is defined by the number of circuit runs that is required to achieve a desired accuracy of the output, scales exponentially to the number of cuts performed.

They can be also understood as techniques for performing a quasiprobability decomposition of quantum channels.
Quasiprobability distribution, which is defined by a set of complex numbers $\{q_i\}$ satisfying $\sum_i q_i = 1$, have recently found a wide range of applications in the area of quantum computing such as error mitigation for NISQ devices \cite{Temme2017, Endo2018} and classical simulation of near-Clifford quantum circuits \cite{Pashayan2015, Howard2017, Bravyi2016, Bennink2017, Seddon2019}.
In particular, Refs. \cite{Bennink2017,Seddon2019,Temme2017,Endo2018} considered a \red{quasiprobability-based simulation of quantum channels;
if a quantum channel $\bm{\Phi}$ can be decomposed as $\bm{\Phi}=\sum_i c_i \bm{\Phi}_i$ where $\bm{\Phi}_i$ and $c_i$ are respectively a channel and a complex coefficient, $\bm{\Phi}$ can be simulated by sampling $\bm{\Phi}_i$ with probability proportional to $|c_i|$ and processing the phase of $c_i$ with classical post-processing.}
The overhead of simulating the channel $\bm{\Phi}$ using this decomposition is quantified by $\sum_i |q_i|$.
If we perform such a decomposition multiple times, the overhead is quantified by the product of $\sum_i |q_i|$, thus leading to an exponential overhead to the number of decomposition performed.
Refs. \cite{Temme2017, Endo2018} have developed techniques to build inverse channels of noise channels using an experimentally available set of quantum gates.
As a technique for a classical simulation, Refs. \cite{Bennink2017, Seddon2019} has considered a quasiprobability decomposition of a non-Clifford channel into Clifford ones.
In this context, we can view the decomposition performed in Ref. \cite{peng2019simulating} as a quasiprobability decomposition of the identity channel into a measurement and state-preparation channel, and one in Ref. \cite{mitarai2019constructing} as a quasiprobability decomposition of a non-local unitary channel into local ones.

In this work, we first define a quantity that we call channel robustness of non-locality in analog to the robustness of magic introduced in Ref. \cite{Howard2017}, which quantifies the minimal possible overhead that can be achieved for quasiprobabilistic simulation of a non-local channel by local channels.
While this quantity is difficult to calculate in general, we show an analytic upper bound for general two-qubit unitary channels by constructing an explicit decomposition, generalizing the technique developed in Ref. \cite{mitarai2019constructing}.
Our previous work \cite{mitarai2019constructing} has only considered decomposition of non-local gates expressed in the form of $e^{i\theta A_1\otimes A_2}$ for Hermitian operators satisfying $A_1^2=I$ and $A_2^2=I$.
In contrast, the decomposition developed in this work performs the cut of a general two-qubit gate in a single-step, leading to a substantially reduced overhead.
Besides the reduced cost, the derivation of the decomposition is delivered more constructively than before which we believe is informative for further optimizations of this approach.
While lower bounds of the defined robustness is also of theoretical interest that can characterize quantumness of a non-local channel, in this work, we focus on upper bounds obtained by explicit decompositions which enable us to actually simulate a nonlocal channel by local channels.
This work develops a theoretical framework for a resource reduction suitable for first-generation quantum devices.

%% file: 30_method.tex
\section{Decomposition of non-local channels into local channels}

\subsection{Notation}

We use the notation $\kket{\rho}$ to express a density matrix $\rho$ to stress that $\rho$ can also be seen as a vector.
Bold-font symbols are to express a quantum channel corresponding to a gate-like operation represented by a normal font.
For example, a unitary channel $\bm{U}$ acts on a state $\kket{\rho}$ as $\bm{U}\kket{\rho} = \kket{U\rho U^\dagger}$ where $U$ is a unitary matrix.
Inner product between two operators $\kket{A}$ and $\kket{B}$ is defined as $\bbraket{A|B}=\Tr(A^\dagger B)$.

\subsection{Channel robustness of non-locality}

In standard eperimental platforms including superconducting qubits and ion traps, it is often thought that the arbitrary single-qubit rotation charactrized by an axis $n=(n_1,n_2,n_3)$ and an angle $\theta$, $R(n,\theta) = \exp\left[-i\theta (\sum_{\alpha}n_\alpha\sigma_\alpha)\right]$, and the single-qubit projective measurements along any axis are somewhat easier operations than two-qubit entangling operations.
Experimentally, the projective measurement is realized by rotating the axis by $R(n,\theta)$ and performing the projective measurement along $z$-axis.
The quantum channel $\bm{M}(n)$ corresponding to the projective measurement is a probabilistic map; when applied to a state $\kket{\rho}$, it returns a state $\bm{\Pi}(\pm n)\kket{\rho}/p_+$ with some probability $p_{\pm}$, where $\Pi(\pm n)$ is a projector to an eigenstate of $\pm \sum_\alpha n_\alpha \sigma_\alpha$ with eigenvalue $+1$.

To implement $\bm{\Pi}(n)$ itself, we can define a probabilistic map $\hat{\bm{\Pi}}(n)$ that takes a state $\kket{\rho}$ to $\bm{\Pi}(n)\kket{\rho}/p_+$ with probability $p_+$ and to $\kket{0}$ with probability $p_-$ where $\kket{0}$ corresponds to the zero matrix.
The map to $\kket{0}$ means simply to ignore the case when the measurement resulted in $-1$.
However, just discarding the $-1$ case is inefficient, especially when we also want to perform $\bm{\Pi}(-n)$.
To resolve this issue, we define a probabilistic map $\tilde{\bm{\Pi}}(n,c_+,c_-)$ that takes a state $\rho$ to $c_\pm\bm{\Pi}(\pm n)\kket{\rho}/p_\pm$ with probability $p_\pm$, where $c_{\pm}\in\{0\}\cup\{e^{i\phi}|\phi\in[0,2\pi]\}$.
\red{Let us define the expected value of a random vector $\kket{\sigma}$ which becomes $\kket{\sigma_i}$ with a probability $p_i$ as $\mathbb{E}[\kket{\sigma}]:=\sum_i p_i \kket{\sigma_i}$.}
Observe that the following holds for any state $\rho$, 
\begin{align}
    \mathbb{E}[\tilde{\bm{\Pi}}(n,c_+,c_-)\kket{\rho}] = c_+\bm{\Pi}(n)\kket{\rho} + c_-\bm{\Pi}(-n)\kket{\rho}.
\end{align}
We write $\mathbb{E}[\tilde{\bm{\Pi}}(n, c_+, c_-)]=c_+\bm{\Pi}(n) + c_-\bm{\Pi}(-n)$ in this sense and henceforth use the notation like this.
$\tilde{\bm{\Pi}}(n,c_+,c_-)$ includes the both of the \red{cases} which we mentioned earlier; if we want to apply only $\bm{\Pi}(n)$ we can set $c_-=0$, and we can also apply both of $\bm{\Pi}(\pm n)$ simultaneously with different coefficients.
The reason we restrict $|c_{\pm}|=1$ is to assure $\left|\Tr[\tilde{\bm{\Pi}}(n,c_+,c_-)\rho]\right|\leq 1$ for any state $\rho$ and any realization of $\tilde{\bm{\Pi}}(n,c_+,c_-)$, thus preventing the decomposition overhead to occur at this stage.

With the above consideration, available local operations in practice, which we denote as $\bm{L}_i$, are the ones that can be written as an arbitrary product of $\bm{R}(n,\theta)$ and $\tilde{\bm{\Pi}}(n)$ and their tensor products.
We denote a set of such possible $\bm{L}_i$ by $\mathcal{L}$.
The most general form of decomposition that we aim to build for a given non-local quantum channel $\bm{\Phi}$ is,
\begin{align}
    \bm{\Phi} = \sum_{i} c_i \bm{L}_i, \label{eq:Phi-decomposition}
\end{align}
where $\bm{L}_i\in \mathcal{L}$.

Given a decomposition above, $\bm{\Phi}$ can be ``simulated'' in a Monte-Carlo manner by sampling $\bm{L}_i$ with probability proportional to $|c_i|$.
More concretely, let us define a probabilistic map $\hat{\bm{\Phi}}$ such that it becomes $\frac{c_i}{|c_i|}\bm{L}_i$ with probability $p_i = |c_i|/W(\bm{\Phi})$ where $W(\bm{\Phi}) = \sum_i |c_i|$.
Then,
\begin{align}
    \mathbb{E}[W(\bm{\Phi})\hat{\bm{\Phi}}] &= W(\bm{\Phi})\times \sum_i \frac{|c_i|}{W(\bm{\Phi})}\frac{c_i}{|c_i|} \bm{L}_i \nonumber \\
    &= \bm{\Phi},
\end{align}
which shows that $W(\bm{\Phi})\hat{\bm{\Phi}}$ becomes equal to $\bm{\Phi}$ when executed for many times.
This algorithm involves only local operations with classical communication (LOCC).
\red{However, note that the above protocol is not a simple probabilistic mixture of LOCC as it multiplies the complex coefficient $c_i/|c_i|$ to each channel $\bm{L}_i$.}

Let us now consider the overhead associated with the decomposition.
In many cases, the output from a quantum system that is evolved with a channel $\bm{\Phi}$ is an expectation value of an observable $O$, which can be written as $\bbra{O}\bm{\Phi}\kket{\rho}$.
$\bbra{O}\bm{\Phi}\kket{\rho}$ is usually estimated by sampling eigenvalues of $O$ from the final state $\bm{\Phi}\kket{\rho}$.
Let the sampled $S$ eigenvalues be $\{o_s\}_{s=1}^S$. 
Normally, we construct an estimator $\widehat{\braket{O}}$ as $\widehat{\braket{O}}=\frac{1}{S}\sum_s o_s$.
Let us assume that absolute value of eigenvalues of $O$ is bounded by $o_{\mathrm{max}}$ and thus $|o_s|\leq o_{\mathrm{max}}$.
Then, by Hoeffding's bound, we can assure that $|\widehat{\braket{O}} - \bbra{O}\bm{V}\kket{\rho}|\leq \epsilon$ with probability at least $1-\delta$ if we take $S=2(o_{\mathrm{max}}/\epsilon)^2 \ln[1/(2\delta)]$.

The number of samples required to achieve the same accuracy increases if one tries to simulate $\bm{\Phi}$ with $\hat{\bm{\Phi}}$.
Since $\mathbb{E}[W(\bm{\Phi})\hat{\bm{\Phi}}]=\bm{\Phi}$, $\mathbb{E}[W(\bm{\Phi})\bbra{O}\hat{\bm{\Phi}}\kket{\rho}]=\bbra{O}\bm{\Phi}\kket{\rho}$
We can construct an estimator $\widehat{\braket{O}}'$ by $\widehat{\braket{O}}' = \frac{1}{S}\sum_s W(\bm{\Phi}) o_s'$ where $o_s'$ is a sample drawn from $\hat{\bm{\Phi}}\kket{\rho}$ with a single realization of $\hat{\bm{\Phi}}$.
The application of $\hat{\bm{\Phi}}$ introduced in the last section involves many stochastic processes; it means to stochastically apply $\bm{L}_i$ with probability $p_i$, and $\bm{L}_i$ itself is a stochastic map involving $\tilde{\bm{\Pi}}(n,c_+,c_-)$.
However, in the end, any realization of $\hat{\bm{\Phi}}$ becomes a single-qubit operation that preserves the magnitude of the trace of $\rho$ or maps the state to $\kket{0}$.
Therefore, it is guaranteed that the absolute value of a sample $o_s'$ obtained by measuring $O$ of $\hat{\bm{\Phi}}\kket{\rho}$ is also bounded by $o_{\mathrm{max}}$.
Again by Hoeffding's bound, $|\widehat{\braket{O}}' - \bbra{O}\bm{\Phi}\kket{\rho}|\leq \epsilon$ with probability at least $1-\delta$ if we take $S=2(W(\bm{\Phi}) o_{\mathrm{max}}/\epsilon)^2 \ln[1/(2\delta)]$.
We can see that $W(\bm{\Phi})^2$ amounts to the overhead of the decomposition.

The above discussion leads us to define the following quantity $\bm{}$ which we call the channel robustness of non-locality,
\begin{align}
    \widetilde{W}(\bm{\Phi}) = \min_{\left\{c_i|\bm{\Phi}=\sum_{i}c_i\bm{L}_i,~\bm{L}_i \in \mathcal{L}\right\}} \sum_i |c_i|. \label{eq:robustness}
\end{align}
$\widetilde{W}(\bm{\Phi})$ quantifies the minimum amount of cost when we perform the simulation of a non-local channel $\bm{\Phi}$ by probabilistic application of the local, experimentally feasible operations.
$\widetilde{W}(\bm{\Phi})$ is submultiplicative, i.e., $\widetilde{W}(\bm{\Phi}_{2}\bm{\Phi}_{1})\leq \widetilde{W}(\bm{\Phi}_2)\widetilde{W}(\bm{\Phi}_1)$, which is proved in Appendix.
This allows us to upper-bound the overhead caused by the decomposition of a chain of quantum channels, $\bm{\Phi}_N\cdots \bm{\Phi}_2\bm{\Phi}_1$ by $\prod_{n=1}^N \widetilde{W}(\bm{\Phi}_n)$.

Note that if we change the available set of operations to some other ones from $\mathcal{L}$,  Eq. (\ref{eq:robustness}) quantifies the overhead of the decomposition in that case.
For example, the overhead of the decomposition of the identity gate presented in Ref. \cite{peng2019simulating} can be quantified by setting the available decomposition to be measure-and-prepare channels.
Another example is the decomposition of non-Clifford circuits into stabilizer-preserving channels considered in Refs. \cite{Bennink2017, Seddon2019}.
The cost for a family of the error mitigation technique called probabilistic error cancellation \cite{Temme2017, Endo2018} is also in relation to this quantity; it is quantified by substituting the target channel $\bm{\Phi}$ with an inverse of a noise channel.

As $\mathcal{L}$ consists of operations with continous parameters, we can also define $\widetilde{W}(\bm{\Phi})$ using a integral instead of a discrete sum. Formally, we can write,
\begin{align}
    \widetilde{W}(\bm{\Phi}) = \min_{\left\{c|\bm{\Phi}=\int c(\lambda)\bm{L}(\lambda) d\lambda,~\bm{L}(\lambda) \in \mathcal{L}\right\}} \int  |c(\lambda)| d\lambda, \label{eq:robustness-continuous}
\end{align}
where $\lambda$ denotes some continuous parameters that specifies an element in $\mathcal{L}$.

The calculation of $\widetilde{W}(\bm{\Phi})$ for a general channel $\bm{\Phi}$ is challenging as it involves a complex minimization procedure.
Nevertheless, in the next section, we give an upper bound of  $\widetilde{W}(\bm{\Phi})$ for a general two-qubit unitary channel $\bm{\Phi}$ by explicitly constructing a decomposition using a complete but not overcomplete basis in $\mathcal{L}$.

\subsection{Upper bound for two-qubit unitary channel}

It is well-known \cite{Kraus2001,Zhang2003} that the non-local part of two-qubit gates can always be written as,
\begin{align}
    U &= \exp\left[i\left(\sum_{\alpha=1}^3 \theta_\alpha \sigma_\alpha\otimes \sigma_\alpha\right)\right]\nonumber \\
    &= \sum_{\alpha=0}^3 u_{\alpha} \sigma_\alpha\otimes\sigma_\alpha \label{eq:two-qubit-unitary},
\end{align}
where $\sigma_0$ is the $2\times 2$ identity operator, and $\sigma_1$, $\sigma_2$ and $\sigma_3$ are Pauli $x$, $y$ and $z$ operators, respectively.
$\theta_\alpha$ is a real parameter, and $u_\alpha$ is a coefficient that is determined from $\{\theta_\alpha\}$.
It leads to the following expression of $\bm{U}$,
\begin{align}
    \bm{U} \kket{\rho} = \sum_{\alpha, \alpha'} u_{\alpha}u_{\alpha'}^*\kket{(\sigma_\alpha\otimes \sigma_\alpha) \rho (\sigma_{\alpha'}\otimes \sigma_{\alpha'})}. \label{eq:two-qubit-unitary-channel}
\end{align}
Note that $\sum_\alpha |u_\alpha|^2 = 1$ follows from the unitarity.

First, we expand the general two-qubit unitary defined in Eq. (\ref{eq:two-qubit-unitary-channel}) using $\kket{\sigma_\beta}$ as a single-qubit basis vector as follows:
\begin{align}
    &\bbra{\sigma_{\beta'} \otimes \sigma_{\gamma'}} \bm{U} \kket{\sigma_\beta \otimes \sigma_\gamma} \nonumber \\
    &= \sum_{\alpha, \alpha' } u_{\alpha}u_{\alpha'}^*\Tr\left[\sigma_{\beta'}\sigma_\alpha\sigma_\beta\sigma_{\alpha'}\right] \Tr \left[ \sigma_{\gamma'} \sigma_\alpha \sigma_\gamma \sigma_{\alpha'} )\right].
\end{align}
From this expression, it is clear that if we can construct a single-qubit channel $\bm{U}_{\alpha\alpha'}$ such that $\bm{U}_{\alpha\alpha'}\rho = \sigma_\alpha\rho\sigma_{\alpha'}$ for any $\rho$, we can write the above as,
\begin{align}
    &\bbra{\sigma_{\beta'} \otimes \sigma_{\gamma'}} \bm{U} \kket{\sigma_\beta \otimes \sigma_\gamma} \nonumber \\
    &= \sum_{\alpha, \alpha'} u_{\alpha}u_{\alpha'}^* \bbra{\sigma_{\beta'}}\bm{U}_{\alpha\alpha'}\kket{\sigma_{\beta}}\bbra{\sigma_{\gamma'}}\bm{U}_{\alpha\alpha'}\kket{\sigma_{\gamma}}\nonumber \\
    &= \sum_{\alpha, \alpha'} u_{\alpha}u_{\alpha'}^* \bbra{\sigma_{\beta'} \otimes \sigma_{\gamma'}} \bm{U}_{\alpha\alpha'}^{\otimes 2} \kket{\sigma_\beta \otimes \sigma_\gamma}.
\end{align}
Therefore, we conclude $\bm{U}=\sum_{\alpha,\alpha'}u_{\alpha}u_{\alpha'}^*\bm{U}_{\alpha\alpha'}^{\otimes 2}$.

Now, we construct $\bm{U}_{\alpha\alpha'}$ with available single-qubit operations.
Observe that,
\begin{align}
    \sigma_\alpha\rho\sigma_{\alpha'} = \frac{1}{2}\left(\sigma_\alpha\rho\sigma_{\alpha'} + \sigma_{\alpha'}\rho\sigma_{\alpha}\right) + \frac{1}{2}\left(\sigma_\alpha\rho\sigma_{\alpha'} - \sigma_{\alpha'}\rho\sigma_{\alpha}\right). 
\end{align}
Let us define the following operators $A_{\alpha\alpha',\pm}$ and $B_{\alpha\alpha',\pm}$ which can be implemented through single-qubit operations:
\begin{align}
    A_{\alpha\alpha',\pm} = \frac{1}{2}\left(\sigma_\alpha \pm \sigma_{\alpha'}\right),\\
    B_{\alpha\alpha',\pm} = \frac{1}{2}\left(\sigma_\alpha \pm i\sigma_{\alpha'}\right).
\end{align}
The corresponding channels $\bm{A}_{\alpha\alpha',\pm}$ and $\bm{B}_{\alpha\alpha',\pm}$ act on a single-qubit density matrix $\rho$ like $A_{\alpha\alpha',\pm}\rho A_{\alpha\alpha',\pm}^\dagger$.
Building on $\bm{A}_{\alpha\alpha',\pm}$ and $\bm{B}_{\alpha\alpha',\pm}$, we further define the following channels:
\begin{align}
    \bm{A}_{\alpha\alpha'} = \bm{A}_{\alpha\alpha',+} - \bm{A}_{\alpha\alpha',-},\\
    \bm{B}_{\alpha\alpha'} = \bm{B}_{\alpha\alpha',+} - \bm{B}_{\alpha\alpha',-}.
\end{align}
With simple algebra, we can see that,
\begin{align}
    \bm{A}_{\alpha\alpha'} \rho = \frac{1}{2}\left(\sigma_\alpha\rho\sigma_{\alpha'} + \sigma_{\alpha'}\rho\sigma_{\alpha}\right), \\
    \bm{B}_{\alpha\alpha'} \rho = \frac{1}{2i}\left(\sigma_\alpha\rho\sigma_{\alpha'} - \sigma_{\alpha'}\rho\sigma_{\alpha}\right).
\end{align}
Therefore, $\bm{U}_{\alpha\alpha'}$ can be written as,
\begin{align}
    \bm{U}_{\alpha\alpha'} = \bm{A}_{\alpha\alpha'} + i\bm{B}_{\alpha\alpha'}.
\end{align}

The above decomposition of $\bm{U}_{\alpha\alpha'}$ leads us to the following decomposition of $\bm{U}$:
\begin{align}
    \bm{U}=\sum_{\alpha\alpha'}u_{\alpha}u_{\alpha'}^*\left(\bm{A}_{\alpha\alpha'} + i\bm{B}_{\alpha\alpha'}\right)^{\otimes 2}.
\end{align}
Note that there are symmetries $\bm{A}_{\alpha\alpha'}=\bm{A}_{\alpha'\alpha}$ and $\bm{B}_{\alpha\alpha'}=-\bm{B}_{\alpha'\alpha}$.
Using them, we rewrite the expression for later convenience,
\begin{align}
    \bm{U}&=\sum_{\alpha}|u_{\alpha}|^2 \bm{\sigma}_\alpha^{\otimes 2} \nonumber\\
    & + \sum_{\alpha<\alpha'} (u_{\alpha}u_{\alpha'}^*+u_{\alpha'}u_{\alpha}^*)\left(\bm{A}_{\alpha\alpha'}^{\otimes 2} - \bm{B}_{\alpha\alpha'}^{\otimes 2} \right)\nonumber \\
    & + \sum_{\alpha<\alpha'} i(u_{\alpha}u_{\alpha'}^*-u_{\alpha'}u_{\alpha}^*)\left(\bm{A}_{\alpha\alpha'}\otimes\bm{B}_{\alpha\alpha'} + \bm{B}_{\alpha\alpha'}\otimes\bm{A}_{\alpha\alpha'}\right). \label{eq:U-decomposition}
\end{align}

To calculate upper bound for $\widetilde{W}(\bm{U})$, we need to formulate Eq. (\ref{eq:U-decomposition}) to fit in the form of Eq. (\ref{eq:Phi-decomposition}).
$\bm{\sigma}_\alpha$, which constitutes the first term of the decomposition, is trivially in $\mathcal{L}$.
Let us now consider $\bm{A}_{\alpha\alpha'}$.
We note that from the symmetry it suffices to consider the case where $\alpha<\alpha'$.
When $\alpha=0$, $\bm{A}_{\alpha\alpha',\pm}$ becomes a projector $\bm{\Pi}(\pm n)$ where $n_{\alpha''} = \delta_{\alpha'\alpha''}$.
Therefore, $\bm{A}_{\alpha\alpha'}$ takes the form of $\tilde{\bm{\Pi}}(n, 1, -1)$, which means $\bm{A}_{0\alpha'}\in \mathcal{L}$.
For $\alpha\neq 0$, $\bm{A}_{\alpha\alpha',\pm}$ is proportional to a single-qubit rotation that swaps the $\alpha$-axis and $\alpha'$-axis.
More concretely, $2\bm{A}_{\alpha\alpha',\pm} \in \mathcal{L}$ for $\alpha\neq 0$ and $\alpha<\alpha'$.
As for $\bm{B}_{\alpha\alpha'}$, when $\alpha=0$, $\bm{B}_{\alpha\alpha',\pm}$ becomes proportional to a single-qubit rotation around $\alpha'$-axis.
Likewise to the previous case, $2\bm{B}_{\alpha\alpha',\pm} \in \bm{L}_i$.
For $\alpha\neq 0$, $\bm{B}_{\alpha\alpha',\pm}$ can be implemented by a projector followed by a flip; for example, $\frac{1}{2}(\sigma_1+i\sigma_2) = \frac{1}{2}\sigma_1(\sigma_0-\sigma_3)$.
With this observation, we can see that the channel $\bm{B}_{\alpha\alpha'}$ in this case can be written as a product of $\tilde{\bm{\Pi}}$ and $\bm{\sigma}_\alpha$ which makes $\bm{B}_{\alpha\alpha'}\in\mathcal{L}$ for $\alpha\neq 0$ and $\alpha<\alpha'$.

Combining the above properties, we can calculate $W(\bm{U}) = \sum_i |c_i|$ for the decomposition given in Eq. (\ref{eq:U-decomposition}) as,
\begin{align}
    W(\bm{U}) = 1 + \sum_{\alpha\neq\alpha'}\left(|u_\alpha u_{\alpha'}^* + u_{\alpha'} u_\alpha^*| + |u_\alpha u_{\alpha'}^* - u_{\alpha'} u_\alpha^*| \right), \label{eq:cost-overhead}
\end{align}
which gives an upper bound of $\widetilde{W}(\bm{U})$.
We note that the operations used in the proposed decomposition, namely $\bm{\sigma}_{\alpha}$ $(\alpha\in\{0,1,2,3\})$, $\bm{A}_{\alpha\alpha'}$ and $\bm{B}_{\alpha\alpha'}$ with $\alpha<\alpha'$ are 16 linearly independent single-qubit channel and thus form a complete basis in the space of single-qubit superoperators.
This means $W(\bm{U})$ is uniquely determined as long as the same basis set is used.

\subsection{Behaviour of \texorpdfstring{$W(\bm{U})$}{TEXT}}

\begin{figure}[]
    \centering
    \includegraphics[width=0.5\linewidth]{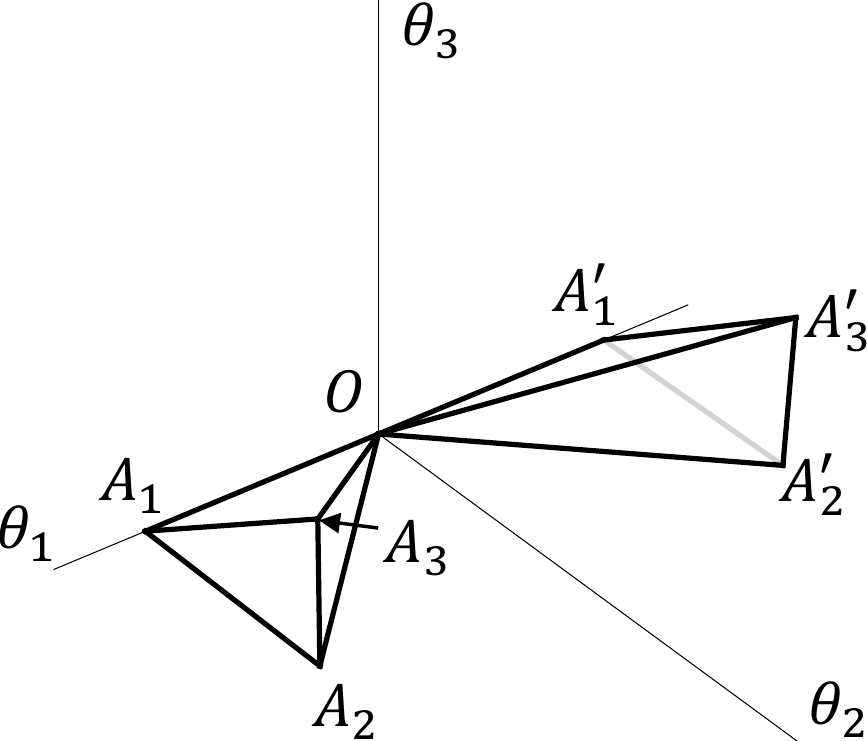}
    \caption{Domain of $(\theta_1,\theta_2,\theta_3)$ in which a two-qubit unitary represented by each point is not locally equivalent to each other. In the figure, $O=(0,0,0)$, $A_1=(\pi/4,0,0)$, $A_2=(\pi/4,\pi/4,0)$, $A_3=(\pi/4,\pi/4,\pi/4)$, $A_1'=(-\pi/4,0,0)$, $A_2'=(-\pi/4,\pi/4,0)$ and $A_3'=(-\pi/4,\pi/4,\pi/4)$.}
    \label{fig:weyl-chamber}
\end{figure}

Here, we numerically investigate the behavior of $W(\bm{U})$ defined in Eq. (\ref{eq:cost-overhead}), restricting the domain of $\{\theta_\alpha\}$ in which each point is not locally equivalent, meaning that a two-qubit unitary represented by a point $(\theta_1, \theta_2, \theta_3)$ cannot be translated to another point in the domain by transforming it with single-qubit unitaries, according to Ref. \cite{Zhang2003}.
In Fig. \ref{fig:weyl-chamber}, we depict such a domain of $\{\theta_\alpha\}$ \footnote{It slightly differs from Ref. \cite{Zhang2003}. We shift half of the tetrahedron presented in Fig. 2 of Ref. \cite{Zhang2003} corresponding to the region $\theta_x \geq \pi/4$ to $\theta_x \leq \pi/0$ using the periodicity of $\theta_x$.}.
Note that there are exceptional local-equivalence in the domain; every point $A_1 A_2 A_3$ and $O A_2 A_3$ is locally equivalent to $A_1'A_2'A_3'$ and $O A_2' A_3'$, respectively.

\begin{figure}[]
    \centering
    \includegraphics[width=0.7\linewidth]{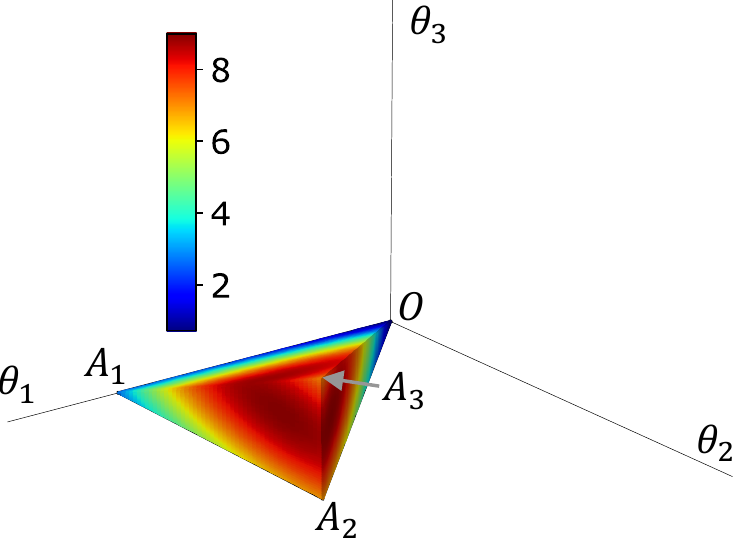}
    \caption{Behaviour of $W(\bm{U})$ on the surface of the tetrahedron $OA_1A_2A_3$.}
    \label{fig:surface-plot}
\end{figure}

Since $W(\bm{U})$ is symmetric to the reflection of $\theta_x$, we only investigate the tetrahedron $OA_1 A_2 A_3$.
In Fig. \ref{fig:surface-plot}, we show the behavior of $W(\bm{U})$ on the surfaces and edges of the domain.
We numerically found that $W(\bm{U})$ is maximized at $(\theta_1, \theta_2, \theta_3)\approx(\pi/4, 0.202\pi, 0.136\pi)$ which lies on the surface $A_1A_2A_3$ with its value being approximately $8.87$.
The behavior of $W(\bm{U})$ seems to be unrelated to other measures such as entangling power of $\bm{U}$ \cite{Kraus2001, Kong2015}; for example, while the point $A_1$ corresponds to controlled-$\sigma_\alpha$ gates which can produce the maximal amount of entanglement and has $W(\bm{U})=3$, $A_3$ which corresponds to the swap gate has $W(\bm{U})=7$.
Although we believe the decomposition given in this work is close to optimal, this counter-intuitive result might be caused by the non-optimality.

%% file: 40_other_approaches.tex
\section{Discussion}

\subsection{Comparison with gate-based decomposition approach}

If we can measure $\braket{\psi_1|\psi_2}$ for some fixed state $\ket{\psi_1}$ and $\ket{\psi_2}$, we can directly utilize the fact that a two-qubit gate is decomposed as $\sum_{\alpha\in \{I,x,y,z\}} u_{\alpha} \sigma_\alpha\otimes\sigma_\alpha$.
\red{As we discuss later, this measurement can be demanding for early days quantum computers.}
Let $V$ be a sequence of gates consisting of alternating layers of single-qubit and two-qubit gates.
Note that any quantum circuit can be written in this form.
$V$ can be written as $V=D_L S_L \cdots D_2 S_2 D_1 S_1$ where $D_i$'s and $S_i$'s are two-qubit and single-qubit gates, respectively.
We assume $D_i = \sum_\alpha d_{\alpha_i} \sigma_{\alpha_i}^{(a_i)} \otimes\sigma_{\alpha_i}^{(b_i)}$ where $\sigma_{\alpha}^{(a)}$ is a Pauli matrix acting on the $a$-th qubit.
Now, focusing on the $i$-th two-qubit gate, we can express an expectation value of an observable $O$ at the end of the circuit as,
\begin{align}\label{eq:gate-based-decomp}
    \braket{0|V^\dagger O V|0} = \sum_{\alpha_i} d_{\alpha'_i}^*d_{\alpha_i} \braket{0|V_{i,\alpha_i'}^\dagger O V_{i,\alpha_i}|0},
\end{align}
where,
\begin{align}
    V_{i,\alpha_i} = D_L S_L\cdots  \sigma_{\alpha_i}^{(a_i)} \otimes\sigma_{\alpha_i}^{(b_i)} \cdots D_2 S_2 D_1 S_1.
\end{align}

This decomposition also allows us to perform a ``virtual'' two-qubit gate on a quantum circuit in the sense that, in $V_{i,\alpha_i}$, the $i$-th two-qubit gate in $V$ is replaced by $\sigma_{\alpha_i}^{(a_i)} \otimes\sigma_{\alpha_i}^{(b_i)}$ which is a tensor product of local operations.
We can do this by the following algorithm.
Let us assume that $O$ is written as $O=\sum_i c_i P_i$, where $P_i$ is a tensor product of Pauli operators.
With this assumption, we can evaluate $\braket{0|V_{i,\alpha_i'}^\dagger O V_{i,\alpha_i}|0}$ by $\sum_k c_k \braket{0|V_{i,\alpha_i'}^\dagger P_k V_{i,\alpha_i}|0}$.
More concretely, we define $\ket{\psi_{i,\alpha_i}} = V_{i,\alpha_i'}\ket{0}$ and $\ket{\psi_{k,i,\alpha_i}}=P_kV_{i,\alpha_i}\ket{\psi_{i,\alpha_i}}$ and then measure $\braket{\psi_{i,\alpha_i'}|\psi_{k,i,\alpha_i}}$ which is possible by the assumption.
If we are to perform the sum of Eq. (\ref{eq:gate-based-decomp}) in a Monte-Carlo manner, we can sample $\alpha'_i$ and $\alpha_i$ with a probability proportional to $|d_{\alpha'_i}^*d_{\alpha'_i}|$.
This leads us to define $G(D_i) :=\sum_{\alpha'_i,\alpha_i}|d_{\alpha'_i}^*d_{\alpha'_i}|$ which quantifies the overhead of the decomposition, that is, we need $G(D_i)^2$ times more samples to reach a desired error compared to the decomposition-free case.

It is trivial that $G(D_i)$ is always smaller than $W(\bm{D}_i)$.
Therefore, if we can measure $\braket{\psi_{i,\alpha_i'}|\psi_{k,i,\alpha_i}}$, it is always better to use this approach.
For example, in a classical simulation we can easily calculate $\braket{\psi_{i,\alpha_i'}|\psi_{k,i,\alpha_i}}$.
However, it is not the case for a quantum computer, in particular for a NISQ device.
Measurement of the overlap $\braket{\psi_{i,\alpha_i'}|\psi_{k,i,\alpha_i}}$, including its phase, is a demanding task.
One way of performing this task is to use a controlled-$V_{i,\alpha_i}$ as mentioned in e.g. Refs. \cite{Bravyi2016,ibe2020calculating}, which is unlikely to be implemented on a NISQ device due to its complexity.
The original motivation of this work and our previous works \cite{mitarai2019constructing, Mitarai2019Methodology} has been to avoid such complex operations.
Note that the famous swap test \cite{Buhrman2001, Garcia-Escartin2013} cannot be applied to this task since it can only evaluate $|\braket{\psi_{i,\alpha_i'}|\psi_{k,i,\alpha_i}}|^2$.
Investigations on the relation between $\widetilde{W}(\bm{D}_i)$ and $G(D_i)$ are left for the furture work.

\subsection{Comparison with the previous work}
In the privous work \cite{mitarai2019constructing}, we have proposed the decomposition for an gate in the form $e^{i\theta A_1\otimes A_2}$ for Hermitian operators $A_1$ and $A_2$ satisfying $A_1^2=I$ and $A_2^2=I$.
It is a special case of this work, which is recovered by setting $u_0 = \cos\theta$ and $u_\alpha = i\sin\theta$ for one chosen $\alpha \in \{1,2,3\}$.
Therefore, the cost overhead of this special case is determined by $1+2|u_0u_\alpha^*-u_\alpha u_0^*|$, which takes maximum at $\theta = \pi/4$.
If we are to decompose a general two-qubit gate in the form of $\exp\left[i\left(\sum_{\alpha=1}^3 \theta_\alpha \sigma_\alpha\otimes \sigma_\alpha\right)\right]$ using this technique, we decompose each of $\exp\left[i\theta_\alpha \sigma_\alpha\otimes \sigma_\alpha\right]$.
Then, the overhead is quantified by the product of $1+2|u_Iu_\alpha^*-u_\alpha u_I|$, which reaches its maximum $3^3=27$ at $\theta_\alpha=\pi/4$ for all $\alpha$.
On the other hand, $W_{\bm{U}}$ defined in Eq. (\ref{eq:cost-overhead}), which quantifies the overhead required by the present approach, becomes $7$, showing substantial improvement. 

While we believe that the decomposition given in this work is close to optimal, there can be better decompositions with smaller $W_{\bm{U}}$.
The search for optimal decomposition will require some form of numerical search.
In the context of classical simulation of near Clifford circuits, Ref. \cite{Bravyi2016} has performed such a search.
However, the optimization of the decomposition considered in this work will be more complicated than the aforementioned work, since the number of available operations is infinitely many as can be seen from Eq. (\ref{eq:robustness-continuous}).
We believe the decomposition proposed in this work can be a good starting point of the optimization if it is not optimal and leave it as future work.

%% file: 70_conclusion.tex
\section{conclusion}

We have introduced a quantity called channel robustness of non-locality which quantifies the minimal amount of overhead required for decomposing non-local channels into local ones with a quasiprobability-based method.
While the calculation of the quantity for general non-local channels is difficult due to the need for a complicated optimization, we have successfully established an upper bound for a general two-qubit unitary channel.
The upper bound is obtained by constructively deriving an explicit decomposition.
Its overhead is substantially lowered compared to the previous work \cite{mitarai2019constructing}.
While we believe the present decomposition is close to optimal, there might be a better decomposition of a general two-qubit channel than the one presented in this work, which we leave as possible future work.
This formalism of decomposing an experimentally challenging channel into a linear combination of experimentally-easy channels allows us to readily perform the decomposition using a quantum device.

\begin{acknowledgments}
    KM is supported by JST PRESTO JPMJPR2019 and KAKENHI No. 20K22330.
    KF is supported by KAKENHI No.16H02211, JST ERATO JPMJER1601, and JST CREST JPMJCR1673.
    This work is supported by MEXT Quantum Leap Flagship Program (MEXT Q-LEAP) Grant Number JPMXS0118067394.
    Program code for generating Fig. \ref{fig:surface-plot} is available at https://github.com/kosukemtr/nonlocal-local-decomposition.
\end{acknowledgments}

%% file: 80_appendix.tex
\section{Submultiplicability of \texorpdfstring{$\widetilde{W}(\bm{\Phi})$}{TEXT}}
\begin{lemma}
    Let $\bm{\Phi}_1$ and $\bm{\Phi}_2$ be any quantum channels and $\bm{\Phi}_{21}=\bm{\Phi}_2\bm{\Phi}_1$. Then, $\widetilde{W}(\bm{\Phi}_{21})\leq \widetilde{W}(\bm{\Phi}_{2})\widetilde{W}(\bm{\Phi}_{1})$.
\end{lemma}
\noindent\textit{proof---}
Let
\begin{align}
    \bm{\Phi}_\mu = \sum_i c_{\mu i}\bm{L}_{\mu i}
\end{align}
and $\sum_i |c_{\mu i}| = \widetilde{W}(\bm{\Phi}_{\mu})$.
Then, $\bm{\Phi}_{21}$ can be decomposed as,
\begin{align}
    \bm{\Phi}_{21} = \sum_{ij} c_{2 i}c_{1 j}\bm{L}_{2 i}\bm{L}_{1 j}.
\end{align}
Because $\bm{L}_{2 i}\bm{L}_{1 j}\in \mathcal{L}$, the above gives a decomposition of $\bm{\Phi}_{21}$ in the form of Eq. \ref{eq:Phi-decomposition}. Therefore,
\begin{align}
    \widetilde{W}(\bm{\Phi}_{21}) &\leq \sum_{ij} |c_{2 i} c_{1 j}| \nonumber \\
    &= \sum_i |c_{2 i}| \sum_j |c_{1 j}| \nonumber \\
    &= \widetilde{W}(\bm{\Phi}_{2})\widetilde{W}(\bm{\Phi}_{1}).
\end{align}
\hfill $\Box$

%% file: 0_main_single_column.bbl
\begin{thebibliography}{25}%
	\makeatletter
	\providecommand \@ifxundefined [1]{%
	 \@ifx{#1\undefined}
	}%
	\providecommand \@ifnum [1]{%
	 \ifnum #1\expandafter \@firstoftwo
	 \else \expandafter \@secondoftwo
	 \fi
	}%
	\providecommand \@ifx [1]{%
	 \ifx #1\expandafter \@firstoftwo
	 \else \expandafter \@secondoftwo
	 \fi
	}%
	\providecommand \natexlab [1]{#1}%
	\providecommand \enquote  [1]{``#1''}%
	\providecommand \bibnamefont  [1]{#1}%
	\providecommand \bibfnamefont [1]{#1}%
	\providecommand \citenamefont [1]{#1}%
	\providecommand \href@noop [0]{\@secondoftwo}%
	\providecommand \href [0]{\begingroup \@sanitize@url \@href}%
	\providecommand \@href[1]{\@@startlink{#1}\@@href}%
	\providecommand \@@href[1]{\endgroup#1\@@endlink}%
	\providecommand \@sanitize@url [0]{\catcode `\\12\catcode `\$12\catcode
	  `\&12\catcode `\#12\catcode `\^12\catcode `\_12\catcode `\%12\relax}%
	\providecommand \@@startlink[1]{}%
	\providecommand \@@endlink[0]{}%
	\providecommand \url  [0]{\begingroup\@sanitize@url \@url }%
	\providecommand \@url [1]{\endgroup\@href {#1}{\urlprefix }}%
	\providecommand \urlprefix  [0]{URL }%
	\providecommand \Eprint [0]{\href }%
	\providecommand \doibase [0]{http://dx.doi.org/}%
	\providecommand \selectlanguage [0]{\@gobble}%
	\providecommand \bibinfo  [0]{\@secondoftwo}%
	\providecommand \bibfield  [0]{\@secondoftwo}%
	\providecommand \translation [1]{[#1]}%
	\providecommand \BibitemOpen [0]{}%
	\providecommand \bibitemStop [0]{}%
	\providecommand \bibitemNoStop [0]{.\EOS\space}%
	\providecommand \EOS [0]{\spacefactor3000\relax}%
	\providecommand \BibitemShut  [1]{\csname bibitem#1\endcsname}%
	\let\auto@bib@innerbib\@empty
	\bibitem [{\citenamefont {Arute}\ \emph {et~al.}(2019)\citenamefont {Arute},
	  \citenamefont {Arya}, \citenamefont {Babbush}, \citenamefont {Bacon},
	  \citenamefont {Bardin}, \citenamefont {Barends}, \citenamefont {Biswas},
	  \citenamefont {Boixo}, \citenamefont {Brandao}, \citenamefont {Buell},
	  \citenamefont {Burkett}, \citenamefont {Chen}, \citenamefont {Chen},
	  \citenamefont {Chiaro}, \citenamefont {Collins}, \citenamefont {Courtney},
	  \citenamefont {Dunsworth}, \citenamefont {Farhi}, \citenamefont {Foxen},
	  \citenamefont {Fowler}, \citenamefont {Gidney}, \citenamefont {Giustina},
	  \citenamefont {Graff}, \citenamefont {Guerin}, \citenamefont {Habegger},
	  \citenamefont {Harrigan}, \citenamefont {Hartmann}, \citenamefont {Ho},
	  \citenamefont {Hoffmann}, \citenamefont {Huang}, \citenamefont {Humble},
	  \citenamefont {Isakov}, \citenamefont {Jeffrey}, \citenamefont {Jiang},
	  \citenamefont {Kafri}, \citenamefont {Kechedzhi}, \citenamefont {Kelly},
	  \citenamefont {Klimov}, \citenamefont {Knysh}, \citenamefont {Korotkov},
	  \citenamefont {Kostritsa}, \citenamefont {Landhuis}, \citenamefont
	  {Lindmark}, \citenamefont {Lucero}, \citenamefont {Lyakh}, \citenamefont
	  {Mandr{\`a}}, \citenamefont {McClean}, \citenamefont {McEwen}, \citenamefont
	  {Megrant}, \citenamefont {Mi}, \citenamefont {Michielsen}, \citenamefont
	  {Mohseni}, \citenamefont {Mutus}, \citenamefont {Naaman}, \citenamefont
	  {Neeley}, \citenamefont {Neill}, \citenamefont {Niu}, \citenamefont {Ostby},
	  \citenamefont {Petukhov}, \citenamefont {Platt}, \citenamefont {Quintana},
	  \citenamefont {Rieffel}, \citenamefont {Roushan}, \citenamefont {Rubin},
	  \citenamefont {Sank}, \citenamefont {Satzinger}, \citenamefont {Smelyanskiy},
	  \citenamefont {Sung}, \citenamefont {Trevithick}, \citenamefont
	  {Vainsencher}, \citenamefont {Villalonga}, \citenamefont {White},
	  \citenamefont {Yao}, \citenamefont {Yeh}, \citenamefont {Zalcman},
	  \citenamefont {Neven},\ and\ \citenamefont {Martinis}}]{Arute2019}%
	  \BibitemOpen
	  \bibfield  {author} {\bibinfo {author} {\bibfnamefont {F.}~\bibnamefont
	  {Arute}}, \bibinfo {author} {\bibfnamefont {K.}~\bibnamefont {Arya}},
	  \bibinfo {author} {\bibfnamefont {R.}~\bibnamefont {Babbush}}, \bibinfo
	  {author} {\bibfnamefont {D.}~\bibnamefont {Bacon}}, \bibinfo {author}
	  {\bibfnamefont {J.~C.}\ \bibnamefont {Bardin}}, \bibinfo {author}
	  {\bibfnamefont {R.}~\bibnamefont {Barends}}, \bibinfo {author} {\bibfnamefont
	  {R.}~\bibnamefont {Biswas}}, \bibinfo {author} {\bibfnamefont
	  {S.}~\bibnamefont {Boixo}}, \bibinfo {author} {\bibfnamefont {F.~G. S.~L.}\
	  \bibnamefont {Brandao}}, \bibinfo {author} {\bibfnamefont {D.~A.}\
	  \bibnamefont {Buell}}, \bibinfo {author} {\bibfnamefont {B.}~\bibnamefont
	  {Burkett}}, \bibinfo {author} {\bibfnamefont {Y.}~\bibnamefont {Chen}},
	  \bibinfo {author} {\bibfnamefont {Z.}~\bibnamefont {Chen}}, \bibinfo {author}
	  {\bibfnamefont {B.}~\bibnamefont {Chiaro}}, \bibinfo {author} {\bibfnamefont
	  {R.}~\bibnamefont {Collins}}, \bibinfo {author} {\bibfnamefont
	  {W.}~\bibnamefont {Courtney}}, \bibinfo {author} {\bibfnamefont
	  {A.}~\bibnamefont {Dunsworth}}, \bibinfo {author} {\bibfnamefont
	  {E.}~\bibnamefont {Farhi}}, \bibinfo {author} {\bibfnamefont
	  {B.}~\bibnamefont {Foxen}}, \bibinfo {author} {\bibfnamefont
	  {A.}~\bibnamefont {Fowler}}, \bibinfo {author} {\bibfnamefont
	  {C.}~\bibnamefont {Gidney}}, \bibinfo {author} {\bibfnamefont
	  {M.}~\bibnamefont {Giustina}}, \bibinfo {author} {\bibfnamefont
	  {R.}~\bibnamefont {Graff}}, \bibinfo {author} {\bibfnamefont
	  {K.}~\bibnamefont {Guerin}}, \bibinfo {author} {\bibfnamefont
	  {S.}~\bibnamefont {Habegger}}, \bibinfo {author} {\bibfnamefont {M.~P.}\
	  \bibnamefont {Harrigan}}, \bibinfo {author} {\bibfnamefont {M.~J.}\
	  \bibnamefont {Hartmann}}, \bibinfo {author} {\bibfnamefont {A.}~\bibnamefont
	  {Ho}}, \bibinfo {author} {\bibfnamefont {M.}~\bibnamefont {Hoffmann}},
	  \bibinfo {author} {\bibfnamefont {T.}~\bibnamefont {Huang}}, \bibinfo
	  {author} {\bibfnamefont {T.~S.}\ \bibnamefont {Humble}}, \bibinfo {author}
	  {\bibfnamefont {S.~V.}\ \bibnamefont {Isakov}}, \bibinfo {author}
	  {\bibfnamefont {E.}~\bibnamefont {Jeffrey}}, \bibinfo {author} {\bibfnamefont
	  {Z.}~\bibnamefont {Jiang}}, \bibinfo {author} {\bibfnamefont
	  {D.}~\bibnamefont {Kafri}}, \bibinfo {author} {\bibfnamefont
	  {K.}~\bibnamefont {Kechedzhi}}, \bibinfo {author} {\bibfnamefont
	  {J.}~\bibnamefont {Kelly}}, \bibinfo {author} {\bibfnamefont {P.~V.}\
	  \bibnamefont {Klimov}}, \bibinfo {author} {\bibfnamefont {S.}~\bibnamefont
	  {Knysh}}, \bibinfo {author} {\bibfnamefont {A.}~\bibnamefont {Korotkov}},
	  \bibinfo {author} {\bibfnamefont {F.}~\bibnamefont {Kostritsa}}, \bibinfo
	  {author} {\bibfnamefont {D.}~\bibnamefont {Landhuis}}, \bibinfo {author}
	  {\bibfnamefont {M.}~\bibnamefont {Lindmark}}, \bibinfo {author}
	  {\bibfnamefont {E.}~\bibnamefont {Lucero}}, \bibinfo {author} {\bibfnamefont
	  {D.}~\bibnamefont {Lyakh}}, \bibinfo {author} {\bibfnamefont
	  {S.}~\bibnamefont {Mandr{\`a}}}, \bibinfo {author} {\bibfnamefont {J.~R.}\
	  \bibnamefont {McClean}}, \bibinfo {author} {\bibfnamefont {M.}~\bibnamefont
	  {McEwen}}, \bibinfo {author} {\bibfnamefont {A.}~\bibnamefont {Megrant}},
	  \bibinfo {author} {\bibfnamefont {X.}~\bibnamefont {Mi}}, \bibinfo {author}
	  {\bibfnamefont {K.}~\bibnamefont {Michielsen}}, \bibinfo {author}
	  {\bibfnamefont {M.}~\bibnamefont {Mohseni}}, \bibinfo {author} {\bibfnamefont
	  {J.}~\bibnamefont {Mutus}}, \bibinfo {author} {\bibfnamefont
	  {O.}~\bibnamefont {Naaman}}, \bibinfo {author} {\bibfnamefont
	  {M.}~\bibnamefont {Neeley}}, \bibinfo {author} {\bibfnamefont
	  {C.}~\bibnamefont {Neill}}, \bibinfo {author} {\bibfnamefont {M.~Y.}\
	  \bibnamefont {Niu}}, \bibinfo {author} {\bibfnamefont {E.}~\bibnamefont
	  {Ostby}}, \bibinfo {author} {\bibfnamefont {A.}~\bibnamefont {Petukhov}},
	  \bibinfo {author} {\bibfnamefont {J.~C.}\ \bibnamefont {Platt}}, \bibinfo
	  {author} {\bibfnamefont {C.}~\bibnamefont {Quintana}}, \bibinfo {author}
	  {\bibfnamefont {E.~G.}\ \bibnamefont {Rieffel}}, \bibinfo {author}
	  {\bibfnamefont {P.}~\bibnamefont {Roushan}}, \bibinfo {author} {\bibfnamefont
	  {N.~C.}\ \bibnamefont {Rubin}}, \bibinfo {author} {\bibfnamefont
	  {D.}~\bibnamefont {Sank}}, \bibinfo {author} {\bibfnamefont {K.~J.}\
	  \bibnamefont {Satzinger}}, \bibinfo {author} {\bibfnamefont {V.}~\bibnamefont
	  {Smelyanskiy}}, \bibinfo {author} {\bibfnamefont {K.~J.}\ \bibnamefont
	  {Sung}}, \bibinfo {author} {\bibfnamefont {M.~D.}\ \bibnamefont
	  {Trevithick}}, \bibinfo {author} {\bibfnamefont {A.}~\bibnamefont
	  {Vainsencher}}, \bibinfo {author} {\bibfnamefont {B.}~\bibnamefont
	  {Villalonga}}, \bibinfo {author} {\bibfnamefont {T.}~\bibnamefont {White}},
	  \bibinfo {author} {\bibfnamefont {Z.~J.}\ \bibnamefont {Yao}}, \bibinfo
	  {author} {\bibfnamefont {P.}~\bibnamefont {Yeh}}, \bibinfo {author}
	  {\bibfnamefont {A.}~\bibnamefont {Zalcman}}, \bibinfo {author} {\bibfnamefont
	  {H.}~\bibnamefont {Neven}}, \ and\ \bibinfo {author} {\bibfnamefont {J.~M.}\
	  \bibnamefont {Martinis}},\ }\href {\doibase 10.1038/s41586-019-1666-5}
	  {\bibfield  {journal} {\bibinfo  {journal} {Nature}\ }\textbf {\bibinfo
	  {volume} {574}},\ \bibinfo {pages} {505} (\bibinfo {year}
	  {2019})}\BibitemShut {NoStop}%
	\bibitem [{\citenamefont {Preskill}(2018)}]{Preskill2018quantumcomputingin}%
	  \BibitemOpen
	  \bibfield  {author} {\bibinfo {author} {\bibfnamefont {J.}~\bibnamefont
	  {Preskill}},\ }\href {\doibase 10.22331/q-2018-08-06-79} {\bibfield
	  {journal} {\bibinfo  {journal} {{Quantum}}\ }\textbf {\bibinfo {volume}
	  {2}},\ \bibinfo {pages} {79} (\bibinfo {year} {2018})}\BibitemShut {NoStop}%
	\bibitem [{\citenamefont {Peruzzo}\ \emph {et~al.}(2014)\citenamefont
	  {Peruzzo}, \citenamefont {McClean}, \citenamefont {Shadbolt}, \citenamefont
	  {Yung}, \citenamefont {Zhou}, \citenamefont {Love}, \citenamefont
	  {Aspuru-Guzik},\ and\ \citenamefont {O'Brien}}]{Peruzzo2014}%
	  \BibitemOpen
	  \bibfield  {author} {\bibinfo {author} {\bibfnamefont {A.}~\bibnamefont
	  {Peruzzo}}, \bibinfo {author} {\bibfnamefont {J.}~\bibnamefont {McClean}},
	  \bibinfo {author} {\bibfnamefont {P.}~\bibnamefont {Shadbolt}}, \bibinfo
	  {author} {\bibfnamefont {M.-H.}\ \bibnamefont {Yung}}, \bibinfo {author}
	  {\bibfnamefont {X.-Q.}\ \bibnamefont {Zhou}}, \bibinfo {author}
	  {\bibfnamefont {P.~J.}\ \bibnamefont {Love}}, \bibinfo {author}
	  {\bibfnamefont {A.}~\bibnamefont {Aspuru-Guzik}}, \ and\ \bibinfo {author}
	  {\bibfnamefont {J.~L.}\ \bibnamefont {O'Brien}},\ }\href
	  {https://doi.org/10.1038/ncomms5213} {\bibfield  {journal} {\bibinfo
	  {journal} {Nature Communications}\ }\textbf {\bibinfo {volume} {5}},\
	  \bibinfo {pages} {4213} (\bibinfo {year} {2014})}\BibitemShut {NoStop}%
	\bibitem [{\citenamefont {McArdle}\ \emph {et~al.}(2020)\citenamefont
	  {McArdle}, \citenamefont {Endo}, \citenamefont {Aspuru-Guzik}, \citenamefont
	  {Benjamin},\ and\ \citenamefont {Yuan}}]{RevModPhys.92.015003}%
	  \BibitemOpen
	  \bibfield  {author} {\bibinfo {author} {\bibfnamefont {S.}~\bibnamefont
	  {McArdle}}, \bibinfo {author} {\bibfnamefont {S.}~\bibnamefont {Endo}},
	  \bibinfo {author} {\bibfnamefont {A.}~\bibnamefont {Aspuru-Guzik}}, \bibinfo
	  {author} {\bibfnamefont {S.~C.}\ \bibnamefont {Benjamin}}, \ and\ \bibinfo
	  {author} {\bibfnamefont {X.}~\bibnamefont {Yuan}},\ }\href {\doibase
	  10.1103/RevModPhys.92.015003} {\bibfield  {journal} {\bibinfo  {journal}
	  {Rev. Mod. Phys.}\ }\textbf {\bibinfo {volume} {92}},\ \bibinfo {pages}
	  {015003} (\bibinfo {year} {2020})}\BibitemShut {NoStop}%
	\bibitem [{\citenamefont {Farhi}\ \emph {et~al.}(2014)\citenamefont {Farhi},
	  \citenamefont {Goldstone},\ and\ \citenamefont {Gutmann}}]{Farhi2014}%
	  \BibitemOpen
	  \bibfield  {author} {\bibinfo {author} {\bibfnamefont {E.}~\bibnamefont
	  {Farhi}}, \bibinfo {author} {\bibfnamefont {J.}~\bibnamefont {Goldstone}}, \
	  and\ \bibinfo {author} {\bibfnamefont {S.}~\bibnamefont {Gutmann}},\
	  }\href@noop {} {\enquote {\bibinfo {title} {A quantum approximate
	  optimization algorithm},}\ } (\bibinfo {year} {2014}),\ \Eprint
	  {http://arxiv.org/abs/1411.4028} {arXiv:1411.4028 [quant-ph]} \BibitemShut
	  {NoStop}%
	\bibitem [{\citenamefont {Mitarai}\ \emph {et~al.}(2018)\citenamefont
	  {Mitarai}, \citenamefont {Negoro}, \citenamefont {Kitagawa},\ and\
	  \citenamefont {Fujii}}]{Mitarai2018}%
	  \BibitemOpen
	  \bibfield  {author} {\bibinfo {author} {\bibfnamefont {K.}~\bibnamefont
	  {Mitarai}}, \bibinfo {author} {\bibfnamefont {M.}~\bibnamefont {Negoro}},
	  \bibinfo {author} {\bibfnamefont {M.}~\bibnamefont {Kitagawa}}, \ and\
	  \bibinfo {author} {\bibfnamefont {K.}~\bibnamefont {Fujii}},\ }\href
	  {\doibase 10.1103/PhysRevA.98.032309} {\bibfield  {journal} {\bibinfo
	  {journal} {Phys. Rev. A}\ }\textbf {\bibinfo {volume} {98}},\ \bibinfo
	  {pages} {032309} (\bibinfo {year} {2018})}\BibitemShut {NoStop}%
	\bibitem [{\citenamefont {Farhi}\ and\ \citenamefont
	  {Neven}(2018)}]{farhi2018classification}%
	  \BibitemOpen
	  \bibfield  {author} {\bibinfo {author} {\bibfnamefont {E.}~\bibnamefont
	  {Farhi}}\ and\ \bibinfo {author} {\bibfnamefont {H.}~\bibnamefont {Neven}},\
	  }\href@noop {} {\enquote {\bibinfo {title} {Classification with quantum
	  neural networks on near term processors},}\ } (\bibinfo {year} {2018}),\
	  \Eprint {http://arxiv.org/abs/1802.06002} {arXiv:1802.06002 [quant-ph]}
	  \BibitemShut {NoStop}%
	\bibitem [{\citenamefont {Bravo-Prieto}\ \emph {et~al.}(2019)\citenamefont
	  {Bravo-Prieto}, \citenamefont {LaRose}, \citenamefont {Cerezo}, \citenamefont
	  {Subasi}, \citenamefont {Cincio},\ and\ \citenamefont
	  {Coles}}]{bravoprieto2019variational}%
	  \BibitemOpen
	  \bibfield  {author} {\bibinfo {author} {\bibfnamefont {C.}~\bibnamefont
	  {Bravo-Prieto}}, \bibinfo {author} {\bibfnamefont {R.}~\bibnamefont
	  {LaRose}}, \bibinfo {author} {\bibfnamefont {M.}~\bibnamefont {Cerezo}},
	  \bibinfo {author} {\bibfnamefont {Y.}~\bibnamefont {Subasi}}, \bibinfo
	  {author} {\bibfnamefont {L.}~\bibnamefont {Cincio}}, \ and\ \bibinfo {author}
	  {\bibfnamefont {P.~J.}\ \bibnamefont {Coles}},\ }\href@noop {} {\enquote
	  {\bibinfo {title} {Variational quantum linear solver},}\ } (\bibinfo {year}
	  {2019}),\ \Eprint {http://arxiv.org/abs/1909.05820} {arXiv:1909.05820
	  [quant-ph]} \BibitemShut {NoStop}%
	\bibitem [{\citenamefont {LaRose}\ \emph {et~al.}(2019)\citenamefont {LaRose},
	  \citenamefont {Tikku}, \citenamefont {O'Neel-Judy}, \citenamefont {Cincio},\
	  and\ \citenamefont {Coles}}]{LaRose_2019}%
	  \BibitemOpen
	  \bibfield  {author} {\bibinfo {author} {\bibfnamefont {R.}~\bibnamefont
	  {LaRose}}, \bibinfo {author} {\bibfnamefont {A.}~\bibnamefont {Tikku}},
	  \bibinfo {author} {\bibfnamefont {E.}~\bibnamefont {O'Neel-Judy}}, \bibinfo
	  {author} {\bibfnamefont {L.}~\bibnamefont {Cincio}}, \ and\ \bibinfo {author}
	  {\bibfnamefont {P.~J.}\ \bibnamefont {Coles}},\ }\href
	  {http://dx.doi.org/10.1038/s41534-019-0167-6} {\bibfield  {journal} {\bibinfo
	   {journal} {npj Quantum Information}\ }\textbf {\bibinfo {volume} {5}},\
	  \bibinfo {pages} {8} (\bibinfo {year} {2019})}\BibitemShut {NoStop}%
	\bibitem [{\citenamefont {Peng}\ \emph {et~al.}(2020)\citenamefont {Peng},
	  \citenamefont {Harrow}, \citenamefont {Ozols},\ and\ \citenamefont
	  {Wu}}]{peng2019simulating}%
	  \BibitemOpen
	  \bibfield  {author} {\bibinfo {author} {\bibfnamefont {T.}~\bibnamefont
	  {Peng}}, \bibinfo {author} {\bibfnamefont {A.~W.}\ \bibnamefont {Harrow}},
	  \bibinfo {author} {\bibfnamefont {M.}~\bibnamefont {Ozols}}, \ and\ \bibinfo
	  {author} {\bibfnamefont {X.}~\bibnamefont {Wu}},\ }\href {\doibase
	  10.1103/PhysRevLett.125.150504} {\bibfield  {journal} {\bibinfo  {journal}
	  {Phys. Rev. Lett.}\ }\textbf {\bibinfo {volume} {125}},\ \bibinfo {pages}
	  {150504} (\bibinfo {year} {2020})}\BibitemShut {NoStop}%
	\bibitem [{\citenamefont {Mitarai}\ and\ \citenamefont
	  {Fujii}(2020)}]{mitarai2019constructing}%
	  \BibitemOpen
	  \bibfield  {author} {\bibinfo {author} {\bibfnamefont {K.}~\bibnamefont
	  {Mitarai}}\ and\ \bibinfo {author} {\bibfnamefont {K.}~\bibnamefont
	  {Fujii}},\ }\href {https://doi.org/10.1088/1367-2630/abd7bc} {\bibfield
	  {journal} {\bibinfo  {journal} {New Journal of Physics}\ ,\ \bibinfo {pages}
	  {accepted}} (\bibinfo {year} {2020})}\BibitemShut {NoStop}%
	\bibitem [{\citenamefont {Temme}\ \emph {et~al.}(2017)\citenamefont {Temme},
	  \citenamefont {Bravyi},\ and\ \citenamefont {Gambetta}}]{Temme2017}%
	  \BibitemOpen
	  \bibfield  {author} {\bibinfo {author} {\bibfnamefont {K.}~\bibnamefont
	  {Temme}}, \bibinfo {author} {\bibfnamefont {S.}~\bibnamefont {Bravyi}}, \
	  and\ \bibinfo {author} {\bibfnamefont {J.~M.}\ \bibnamefont {Gambetta}},\
	  }\href {\doibase 10.1103/PhysRevLett.119.180509} {\bibfield  {journal}
	  {\bibinfo  {journal} {Phys. Rev. Lett.}\ }\textbf {\bibinfo {volume} {119}},\
	  \bibinfo {pages} {180509} (\bibinfo {year} {2017})}\BibitemShut {NoStop}%
	\bibitem [{\citenamefont {Endo}\ \emph {et~al.}(2018)\citenamefont {Endo},
	  \citenamefont {Benjamin},\ and\ \citenamefont {Li}}]{Endo2018}%
	  \BibitemOpen
	  \bibfield  {author} {\bibinfo {author} {\bibfnamefont {S.}~\bibnamefont
	  {Endo}}, \bibinfo {author} {\bibfnamefont {S.~C.}\ \bibnamefont {Benjamin}},
	  \ and\ \bibinfo {author} {\bibfnamefont {Y.}~\bibnamefont {Li}},\ }\href
	  {\doibase 10.1103/PhysRevX.8.031027} {\bibfield  {journal} {\bibinfo
	  {journal} {Phys. Rev. X}\ }\textbf {\bibinfo {volume} {8}},\ \bibinfo {pages}
	  {031027} (\bibinfo {year} {2018})}\BibitemShut {NoStop}%
	\bibitem [{\citenamefont {Pashayan}\ \emph {et~al.}(2015)\citenamefont
	  {Pashayan}, \citenamefont {Wallman},\ and\ \citenamefont
	  {Bartlett}}]{Pashayan2015}%
	  \BibitemOpen
	  \bibfield  {author} {\bibinfo {author} {\bibfnamefont {H.}~\bibnamefont
	  {Pashayan}}, \bibinfo {author} {\bibfnamefont {J.~J.}\ \bibnamefont
	  {Wallman}}, \ and\ \bibinfo {author} {\bibfnamefont {S.~D.}\ \bibnamefont
	  {Bartlett}},\ }\href {\doibase 10.1103/PhysRevLett.115.070501} {\bibfield
	  {journal} {\bibinfo  {journal} {Phys. Rev. Lett.}\ }\textbf {\bibinfo
	  {volume} {115}},\ \bibinfo {pages} {070501} (\bibinfo {year}
	  {2015})}\BibitemShut {NoStop}%
	\bibitem [{\citenamefont {Howard}\ and\ \citenamefont
	  {Campbell}(2017)}]{Howard2017}%
	  \BibitemOpen
	  \bibfield  {author} {\bibinfo {author} {\bibfnamefont {M.}~\bibnamefont
	  {Howard}}\ and\ \bibinfo {author} {\bibfnamefont {E.}~\bibnamefont
	  {Campbell}},\ }\href {\doibase 10.1103/PhysRevLett.118.090501} {\bibfield
	  {journal} {\bibinfo  {journal} {Phys. Rev. Lett.}\ }\textbf {\bibinfo
	  {volume} {118}},\ \bibinfo {pages} {090501} (\bibinfo {year}
	  {2017})}\BibitemShut {NoStop}%
	\bibitem [{\citenamefont {Bravyi}\ \emph {et~al.}(2016)\citenamefont {Bravyi},
	  \citenamefont {Smith},\ and\ \citenamefont {Smolin}}]{Bravyi2016}%
	  \BibitemOpen
	  \bibfield  {author} {\bibinfo {author} {\bibfnamefont {S.}~\bibnamefont
	  {Bravyi}}, \bibinfo {author} {\bibfnamefont {G.}~\bibnamefont {Smith}}, \
	  and\ \bibinfo {author} {\bibfnamefont {J.~A.}\ \bibnamefont {Smolin}},\
	  }\href {\doibase 10.1103/PhysRevX.6.021043} {\bibfield  {journal} {\bibinfo
	  {journal} {Phys. Rev. X}\ }\textbf {\bibinfo {volume} {6}},\ \bibinfo {pages}
	  {021043} (\bibinfo {year} {2016})}\BibitemShut {NoStop}%
	\bibitem [{\citenamefont {Bennink}\ \emph {et~al.}(2017)\citenamefont
	  {Bennink}, \citenamefont {Ferragut}, \citenamefont {Humble}, \citenamefont
	  {Laska}, \citenamefont {Nutaro}, \citenamefont {Pleszkoch},\ and\
	  \citenamefont {Pooser}}]{Bennink2017}%
	  \BibitemOpen
	  \bibfield  {author} {\bibinfo {author} {\bibfnamefont {R.~S.}\ \bibnamefont
	  {Bennink}}, \bibinfo {author} {\bibfnamefont {E.~M.}\ \bibnamefont
	  {Ferragut}}, \bibinfo {author} {\bibfnamefont {T.~S.}\ \bibnamefont
	  {Humble}}, \bibinfo {author} {\bibfnamefont {J.~A.}\ \bibnamefont {Laska}},
	  \bibinfo {author} {\bibfnamefont {J.~J.}\ \bibnamefont {Nutaro}}, \bibinfo
	  {author} {\bibfnamefont {M.~G.}\ \bibnamefont {Pleszkoch}}, \ and\ \bibinfo
	  {author} {\bibfnamefont {R.~C.}\ \bibnamefont {Pooser}},\ }\href {\doibase
	  10.1103/PhysRevA.95.062337} {\bibfield  {journal} {\bibinfo  {journal} {Phys.
	  Rev. A}\ }\textbf {\bibinfo {volume} {95}},\ \bibinfo {pages} {062337}
	  (\bibinfo {year} {2017})}\BibitemShut {NoStop}%
	\bibitem [{\citenamefont {Seddon}\ and\ \citenamefont
	  {Campbell}(2019)}]{Seddon2019}%
	  \BibitemOpen
	  \bibfield  {author} {\bibinfo {author} {\bibfnamefont {J.~R.}\ \bibnamefont
	  {Seddon}}\ and\ \bibinfo {author} {\bibfnamefont {E.~T.}\ \bibnamefont
	  {Campbell}},\ }\href {\doibase 10.1098/rspa.2019.0251} {\bibfield  {journal}
	  {\bibinfo  {journal} {Proceedings of the Royal Society A: Mathematical,
	  Physical and Engineering Sciences}\ }\textbf {\bibinfo {volume} {475}},\
	  \bibinfo {pages} {20190251} (\bibinfo {year} {2019})}\BibitemShut {NoStop}%
	\bibitem [{\citenamefont {Kraus}\ and\ \citenamefont
	  {Cirac}(2001)}]{Kraus2001}%
	  \BibitemOpen
	  \bibfield  {author} {\bibinfo {author} {\bibfnamefont {B.}~\bibnamefont
	  {Kraus}}\ and\ \bibinfo {author} {\bibfnamefont {J.~I.}\ \bibnamefont
	  {Cirac}},\ }\href {\doibase 10.1103/PhysRevA.63.062309} {\bibfield  {journal}
	  {\bibinfo  {journal} {Phys. Rev. A}\ }\textbf {\bibinfo {volume} {63}},\
	  \bibinfo {pages} {062309} (\bibinfo {year} {2001})}\BibitemShut {NoStop}%
	\bibitem [{\citenamefont {Zhang}\ \emph {et~al.}(2003)\citenamefont {Zhang},
	  \citenamefont {Vala}, \citenamefont {Sastry},\ and\ \citenamefont
	  {Whaley}}]{Zhang2003}%
	  \BibitemOpen
	  \bibfield  {author} {\bibinfo {author} {\bibfnamefont {J.}~\bibnamefont
	  {Zhang}}, \bibinfo {author} {\bibfnamefont {J.}~\bibnamefont {Vala}},
	  \bibinfo {author} {\bibfnamefont {S.}~\bibnamefont {Sastry}}, \ and\ \bibinfo
	  {author} {\bibfnamefont {K.~B.}\ \bibnamefont {Whaley}},\ }\href {\doibase
	  10.1103/PhysRevA.67.042313} {\bibfield  {journal} {\bibinfo  {journal} {Phys.
	  Rev. A}\ }\textbf {\bibinfo {volume} {67}},\ \bibinfo {pages} {042313}
	  (\bibinfo {year} {2003})}\BibitemShut {NoStop}%
	\bibitem [{\citenamefont {Kong}\ \emph {et~al.}(2015)\citenamefont {Kong},
	  \citenamefont {Zhao}, \citenamefont {Yang},\ and\ \citenamefont
	  {Cao}}]{Kong2015}%
	  \BibitemOpen
	  \bibfield  {author} {\bibinfo {author} {\bibfnamefont {F.-Z.}\ \bibnamefont
	  {Kong}}, \bibinfo {author} {\bibfnamefont {J.-L.}\ \bibnamefont {Zhao}},
	  \bibinfo {author} {\bibfnamefont {M.}~\bibnamefont {Yang}}, \ and\ \bibinfo
	  {author} {\bibfnamefont {Z.-L.}\ \bibnamefont {Cao}},\ }\href {\doibase
	  10.1103/PhysRevA.92.012127} {\bibfield  {journal} {\bibinfo  {journal} {Phys.
	  Rev. A}\ }\textbf {\bibinfo {volume} {92}},\ \bibinfo {pages} {012127}
	  (\bibinfo {year} {2015})}\BibitemShut {NoStop}%
	\bibitem [{\citenamefont {Ibe}\ \emph {et~al.}(2020)\citenamefont {Ibe},
	  \citenamefont {Nakagawa}, \citenamefont {Yamamoto}, \citenamefont {Mitarai},
	  \citenamefont {Gao},\ and\ \citenamefont {Kobayashi}}]{ibe2020calculating}%
	  \BibitemOpen
	  \bibfield  {author} {\bibinfo {author} {\bibfnamefont {Y.}~\bibnamefont
	  {Ibe}}, \bibinfo {author} {\bibfnamefont {Y.~O.}\ \bibnamefont {Nakagawa}},
	  \bibinfo {author} {\bibfnamefont {T.}~\bibnamefont {Yamamoto}}, \bibinfo
	  {author} {\bibfnamefont {K.}~\bibnamefont {Mitarai}}, \bibinfo {author}
	  {\bibfnamefont {Q.}~\bibnamefont {Gao}}, \ and\ \bibinfo {author}
	  {\bibfnamefont {T.}~\bibnamefont {Kobayashi}},\ }\href@noop {} {\enquote
	  {\bibinfo {title} {Calculating transition amplitudes by variational quantum
	  eigensolvers},}\ } (\bibinfo {year} {2020}),\ \Eprint
	  {http://arxiv.org/abs/2002.11724} {arXiv:2002.11724 [quant-ph]} \BibitemShut
	  {NoStop}%
	\bibitem [{\citenamefont {Mitarai}\ and\ \citenamefont
	  {Fujii}(2019)}]{Mitarai2019Methodology}%
	  \BibitemOpen
	  \bibfield  {author} {\bibinfo {author} {\bibfnamefont {K.}~\bibnamefont
	  {Mitarai}}\ and\ \bibinfo {author} {\bibfnamefont {K.}~\bibnamefont
	  {Fujii}},\ }\href {\doibase 10.1103/PhysRevResearch.1.013006} {\bibfield
	  {journal} {\bibinfo  {journal} {Phys. Rev. Research}\ }\textbf {\bibinfo
	  {volume} {1}},\ \bibinfo {pages} {013006} (\bibinfo {year}
	  {2019})}\BibitemShut {NoStop}%
	\bibitem [{\citenamefont {Buhrman}\ \emph {et~al.}(2001)\citenamefont
	  {Buhrman}, \citenamefont {Cleve}, \citenamefont {Watrous},\ and\
	  \citenamefont {de~Wolf}}]{Buhrman2001}%
	  \BibitemOpen
	  \bibfield  {author} {\bibinfo {author} {\bibfnamefont {H.}~\bibnamefont
	  {Buhrman}}, \bibinfo {author} {\bibfnamefont {R.}~\bibnamefont {Cleve}},
	  \bibinfo {author} {\bibfnamefont {J.}~\bibnamefont {Watrous}}, \ and\
	  \bibinfo {author} {\bibfnamefont {R.}~\bibnamefont {de~Wolf}},\ }\href
	  {\doibase 10.1103/PhysRevLett.87.167902} {\bibfield  {journal} {\bibinfo
	  {journal} {Phys. Rev. Lett.}\ }\textbf {\bibinfo {volume} {87}},\ \bibinfo
	  {pages} {167902} (\bibinfo {year} {2001})}\BibitemShut {NoStop}%
	\bibitem [{\citenamefont {Garcia-Escartin}\ and\ \citenamefont
	  {Chamorro-Posada}(2013)}]{Garcia-Escartin2013}%
	  \BibitemOpen
	  \bibfield  {author} {\bibinfo {author} {\bibfnamefont {J.~C.}\ \bibnamefont
	  {Garcia-Escartin}}\ and\ \bibinfo {author} {\bibfnamefont {P.}~\bibnamefont
	  {Chamorro-Posada}},\ }\href {\doibase 10.1103/PhysRevA.87.052330} {\bibfield
	  {journal} {\bibinfo  {journal} {Phys. Rev. A}\ }\textbf {\bibinfo {volume}
	  {87}},\ \bibinfo {pages} {052330} (\bibinfo {year} {2013})}\BibitemShut
	  {NoStop}%
	\end{thebibliography}
